\documentclass{aa}  

\usepackage{graphicx}
\usepackage{txfonts}
\usepackage{natbib}
\def \inte {\emph{INTEGRAL}}
\def \xmm {\emph{XMM-Newton}}
\def \sax {\emph{BeppoSAX}}
\def \sw {{\it Swift}}

\def \rxte {\emph{RXTE}}
\def \src {\mbox{IGR~J18483$-$0311}}

\begin{document}

   \title{Spectral and temporal properties of the supergiant fast X-ray transient IGR~J18483--0311 
          observed by INTEGRAL}
 \titlerunning{INTEGRAL observations of the SFXT IGR~J18483--0311}

   \author{L. Ducci
          \inst{1}
          \and
          V. Doroshenko\inst{1}
          \and
          M. Sasaki\inst{1}
          \and
          A. Santangelo\inst{1}
          \and
          P. Esposito\inst{2}
          \and
          P. Romano\inst{3}
          \and
          S. Vercellone\inst{3}
          }

   \institute{Institut f\"ur Astronomie und Astrophysik, Eberhard Karls Universit\"at, 
              Sand 1, 72076 T\"ubingen, Germany\\
              \email{ducci@astro.uni-tuebingen.de}
              \and
              INAF, Istituto di Astrofisica Spaziale e Fisica Cosmica - Milano, 
              Via E.\ Bassini 15,   I-20133 Milano,  Italy 
              \and
              INAF, Istituto di Astrofisica Spaziale e Fisica Cosmica - Palermo,
              Via U.\ La Malfa 153, I-90146 Palermo, Italy
             }

   \date{}

 
  \abstract
   {\src\ is a supergiant fast X-ray transient whose compact object is located 
    in a wide (18.5 d) and eccentric ($e \sim0.4$) orbit, which shows sporadic outbursts 
    that reach X-ray luminosities of $\sim$10$^{36}$~erg~s$^{-1}$.}
   {We investigated the timing properties of \src\ and studied the spectra
    during bright outbursts by fitting physical models based on thermal and bulk
    Comptonization processes for accreting compact objects.}
   {We analysed archival \inte\ data collected in the period 2003--2010, focusing
    on the observations with \src\ in outburst.
    We searched for pulsations in the \inte\ light curves of each outburst.
    We took advantage of the broadband 
    observing capability of \inte\ for the spectral analysis.}
   {We observed 15 outbursts, seven of which we report here for the first time.
    This data analysis almost doubles the statistics of flares of this binary system
    detected by \inte.
    A refined timing analysis did not reveal a significant periodicity
    in the \inte\ observation where a $\sim$21~s pulsation was previously detected.
    Neither did we find evidence for pulsations in the X-ray light curve
    of an archival \xmm\ observation of \src.
    In the light of these results the nature of the compact object in \src\ is unclear.
    The broadband X-ray spectrum of \src\ in outburst is well fitted 
    by a thermal and bulk Comptonization
    model of blackbody seed photons by the infalling material in the accretion column of a neutron star.
    We also obtained a new measurement of the orbital period using the \sw/BAT light curve.}
   {}

   \keywords{X-rays: binaries -- stars: individual: IGR J18483-0311 - stars: neutron -- X-rays: stars}

   \maketitle
%

\section{Introduction}
\label{sect. introduction}

In the past decade a new generation of sensitive hard X-ray telescopes
led to the discovery of a new class of high-mass X-ray binaries (HMXBs)
with supergiant companions that are called supergiant fast X-ray transients 
(SFXTs; \citealt{Smith04}; \citealt{Sguera05}; \citealt{Negueruela06}).
The large majority of the members of this class were discovered thanks
to the monitoring of the Galactic plane performed by \inte\ 
(INTErnational Gamma-Ray Astrophysics Laboratory).
SFXTs show sporadic outbursts lasting a few days composed of flares
lasting from minutes to hours (\citealt{Romano07}; \citealt{Sidoli09}).
SFXTs are also characterized by a wide dynamic range, spanning three to five
orders of magnitude, from a quiescent luminosity of $\approx 10^{32}$~erg~s$^{-1}$
to the outburst peak luminosity of $10^{36}-10^{37}$~erg~s$^{-1}$ \citep{intZand05}.
The X-ray spectra of this class can be fitted with models
used to describe the X-ray emission from HMXBs that host pulsars
(e.g. \citealt{Walter07}).
Currently, there are ten confirmed SFXTs, five of which have 
a detected periodic signal that has been interpreted as the spin period of an
X-ray pulsar (see e.g. \citealt{Sidoli13}).
Different accretion mechanisms have been proposed to explain the properties of SFXTs.
These involve accretion from an inhomogeneous and anisotropic wind
(see e.g. \citealt{intZand05}; \citealt{Sidoli07}; \citealt{Negueruela08}),
gating mechanisms 
(\citealt{Grebenev07}; \citealt{Bozzo08}),
and intermittent accretion flow onto a neutron star
(e.g. \citealt{Lamb77}; \citealt{Ducci10}).

\src\ is an SFXT 
discovered in 2003 April by \citet{Chernyakova03}
during an observation of the Galactic Centre field with
the IBIS/ISGRI instrument onboard the \inte\ satellite.
These authors observed a transient X-ray emission with a flux of 
$\sim$10~mCrab in the $15-40$~keV band that suddenly increased
to $\sim$40~mCrab  during a $\sim$30 minutes flare on 2003 April 26.
After its discovery, \citet{Molkov03,Molkov04} reported
other \inte\ detections of \src\ in 2003 March-May,
while \citet{intZand04} reported earlier detections
of \src\ with the \sax\ Wide Field Cameras in 1997 April 26
and 1996 July-December.
\citet{Chernyakova03} suggested a possible association
with the \emph{ROSAT}/HRI source 1RXH~J184817.3-031017.
Two optical sources in the USNO-B1 catalogue \citep{Monet03},
one of which was also detected at near-infrared wavelengths in the 
2MASS catalogue \citep{Skrutskie06},
were found within the X-ray positional uncertainty of the HRI instrument
\citep{Stephen06}.
No significant radio emission was found
within the error circle of \src\ \citep{Pandey06}.

\citet{Levine06} detected a periodicity of $\sim$18.55~d
in the \emph{Rossi X-ray Timing Explorer} (\rxte)
All-Sky Monitor (ASM) X-ray light curve,
interpreted as the orbital period of \src\ \citep{Rahoui08}.
This measurement was refined by \citet{Levine11} to $18.545 \pm 0.003$~d
by using the \rxte/ASM light curve.

\citet{Sguera07} reported the observation of five new outbursts
and a periodicity of $21.0526 \pm 0.0005$~s in the $4-20$~keV \inte/JEM-X
light curve during the outburst of 2006 April 19.
They also searched for periodicities with the IBIS/ISGRI instrument
onboard the \inte\ satellite, but no pulsations were detected. 
\citet{Sguera07} attributed the $\sim$21~s periodicity to the spin period 
of an X-ray pulsar.
\citet{Giunta09} observed \src\ with \xmm\ during a low-luminosity state
($L_{\rm 0.5-10\,keV} \approx 10^{34}$ erg s$^{-1}$).
They searched for the periodicity reported by \citet{Sguera07}
by computing a power spectrum of the \xmm\ data,
but did not find significant evidence for pulsations.
Then, they searched for it in a small window centred on $21.0526$~s
by performing Rayleigh ($Z^2_1$) analysis \citep{Buccheri83},
and reported the detection of a periodicity 
at $21.033 \pm 0.004$~s as the most significant peak
within the examined range.

\citet{Sguera07} fitted the broad-band joint \inte\ spectrum (JEM-X + IBIS/ISGRI, $3-50$~keV)
of \src\ during the outburst in 2006 April 19 
with an absorbed cutoff power-law with $\Gamma = 1.4 \pm 0.3$,
$E_c = 22{+8 \atop -5}$~keV, and $N_{\rm H} = 9{+5 \atop -4}\times 10^{22}$ cm$^{-2}$.
Based on the X-ray spectral shape, the detected periodicities,
the high intrinsic absorption, and the highly reddened optical counterpart,
\citet{Sguera07} proposed that \src\ is an HMXB with a neutron star.

\citet{Chaty08} and 
\citet{Rahoui08} performed optical and near-infrared photometry and spectroscopy
of the donor star of \src,
showing that it is a B0.5a supergiant.
The nature of the donor star together with the fast X-ray transient
activity that characterizes \src\ led \citet{Rahoui08} to the conclusion
that the source is an SFXT.
The spectral classification of the donor star
was more recently refined by \citet{Torrejon10}.
They found that the donor star is a B0.5-B1~Iab
located at $\sim$2.8~kpc.

\citet{Romano10} reported the results from a \sw\ monitoring of \src\
lasting $\sim$28~d, consecutively covering more than one orbital period.
The orbital modulation of the \sw/XRT light curve was interpreted
as wind accretion along a highly eccentric orbit \citep{Romano10}.
The dynamical range of \src\ observed during the \sw/XRT monitoring
was wider than $1200$.
Superimposed on the long-term orbital modulation, \src\
showed the typical variability of SFXTs on time scales of a few hours.
\citet{Romano10} interpreted the flaring activity of \src\ in terms
of the accretion of an inhomogeneous wind, using the clumpy-wind model
of \citet{Ducci09}. \citet{Grebenev09} proposed another interpretation 
based on gating mechanisms, where magnetic and centrifugal barriers
(which depend on the spin period and magnetic field of the neutron star) 
influence the accretion rate, producing the observed flaring activity.

%
\sw, \xmm, and \inte\ observations of \src\ spanning luminosities from $\sim$10$^{33}$ erg s$^{-1}$
to $\sim$10$^{36}$ erg s$^{-1}$ did not show an appreciable change in the spectral shape
(\citealt{Romano10}; \citealt{Giunta09}; \citealt{Sguera07}).

In this paper we present a spectral and timing analysis of the SFXT
\src. We used archival \inte\ data collected during the period 2003--2010
(corresponding to a total exposure time of $\sim$381~ks)
with emphasis on
the \inte\ pointings showing \src\ in outburst.
Taking advantage of the high broadband observing capability
of \inte, we applied phenomenological spectral models
and, for the first time for this source, physical models
typically used to describe the X-ray emission from accreting 
compact objects.
We also performed a timing analysis of the \inte\ light curves of \src\
to search for periodic signals. 
For this purpose, we also reanalysed an archival \xmm\ observation.
The observations and data analysis methods are described in Sect. \ref{sect. data analysis}.
We report on our results and discuss them 
in Sects. \ref{sect. timing analysis} and \ref{sect. spectral analysis}.

\begin{table}
\caption{Outbursts of \src\ analysed in this paper. The orbital phase locations 
         $\Delta \phi$ were calculated using our measurements of the orbital period 
         and zero phase ephemeris (see Sect. \ref{sect. orbital modulation}).}             
\label{table:list outbursts}      
\begin{center}   
\resizebox{\columnwidth}{!}{     
\begin{tabular}{lcccc} 
\hline\hline  
\noalign{\smallskip}   
  No.      &      Obs. date    &  $\Delta \phi$  &  Orbital   & $T_{\rm exp}$ \\ 
           &        (MJD)      &                 & revolution &    (ks)     \\
\noalign{\smallskip}
\hline
{ }\,1$^a$ & 52735.2$-$52735.5 &   $0.36-0.38$   &   0058     & { }7.5     \\
\noalign{\smallskip}
{ }\,2$^b$ & 52755.2$-$52755.4 &   $0.44-0.45$   &   0065     & { }3.6     \\
\noalign{\smallskip}
{ }\,3$^c$ & 52770.9$-$52772.1 &   $0.28-0.35$   &   0070     &   29.0     \\
\noalign{\smallskip}
{ }\,4$^c$ & 53082.9$-$53083.4 &   $0.10-0.13$   &   0174     &   10.0     \\
\noalign{\smallskip}
{ }\,5$^c$ & 53121.7$-$53122.6 &   $0.19-0.24$   & 0187, 0188 &   91.3     \\
\noalign{\smallskip}
{ }\,6$^c$ & 53253.2$-$53253.6 &   $0.28-0.30$   &   0231     &   13.5     \\
\noalign{\smallskip}
{ }\,7$^d$ & 53663.2$-$53663.8 &   $0.38-0.41$   &   0368     &   20.8     \\
\noalign{\smallskip}
{ }\,8$^c$ & 53844.6$-$53846.0 &   $0.16-0.23$   &   0429     &   51.1     \\
\noalign{\smallskip}
{ }\,9$^e$ & 54364.4$-$54365.2 &   $0.17-0.22$   &   0603     &   46.4     \\
\noalign{\smallskip}
    10$^e$ & 54924.6$-$54925.0 &   $0.37-0.39$   &   0790     &   18.1     \\
\noalign{\smallskip}
    11$^e$ & 54926.5$-$54928.2 &   $0.47-0.56$   &   0791     &   21.5     \\
\noalign{\smallskip}
    12$^e$ & 55109.1$-$55109.3 &   $0.31-0.32$   &   0852     &   15.8     \\
\noalign{\smallskip}
    13$^e$ & 55274.3$-$55274.8 &   $0.22-0.24$   &   0907     &   17.5     \\
\noalign{\smallskip}
    14$^e$ & 55280.2$-$55281.8 &   $0.53-0.62$   &   0909     &   14.2     \\
\noalign{\smallskip}  
    15$^e$ & 55284.3$-$55284.9 &   $0.75-0.79$   &   0910     &   19.8     \\
\noalign{\smallskip}
\hline                         
\end{tabular}
}
\end{center}
Notes:\\
$^a$: \citet{Molkov03}\\
$^b$: \citet{Chernyakova03}\\
$^c$: \citet{Sguera07}\\
$^d$: \citet{Ducci10}\\
$^e$: this work\\
\end{table}

\section{Observations and data analysis}
\label{sect. data analysis}

\subsection{INTEGRAL} \label{sect. integral}

The ESA satellite \inte,
launched on 2002 October 17 \citep{Winkler03}, carries 
three X-ray and gamma-ray instruments.
In this paper we used the imager IBIS (Imager on Board \inte\ Satellite, 
\citealt{Ubertini03}), and the X-ray monitors JEM-X 
(Joint European Monitor for X-ray; \citealt{Lund03}),
which are co-aligned with overlapping fields of view 
and operate simultaneously with IBIS.

IBIS is a coded-mask telescope composed of two detector arrays: ISGRI 
(\inte\ Soft Gamma-Ray Imager, \citealt{Lebrun03}), operating in the 15$-$400~keV band,
and PICsIT (Pixellated Imaging CaeSium Iodide Telescope, \citealt{Dicocco03}),
operating in the 180$-$2000~keV band. IBIS has a fully coded field of view of $9^\circ \times 9^\circ$,
and a partially coded field of view of $29^\circ \times 29^\circ$.

JEM-X consists of two coaligned coded mask telescopes, JEM-X1 and JEM-X2,
which are switched on alternately during the mission. 
JEM-X allows imaging, spectral, and timing analyses in the energy range of $3-35$~keV.
The fully illuminated JEM-X field of view has a diameter of $4.8^\circ$.

\inte\ orbits Earth with a highly eccentric orbit with a 
revolution period of about 72 hours. Each orbital revolution
consists of pointings of about 2000~s called science windows (ScWs).

We performed the reduction and analysis of IBIS/ISGRI and JEM-X data 
using the Off-line Science Analysis (OSA) 10.0 software \citep{Goldwurm03}.
We analysed the public data between 2003 January and 2010 December
where \src\ was in the IBIS/ISGRI field of view.
We only considered ScWs where \src\ was detected with at least $5\sigma$
significance in the energy 18$-$50~keV (corresponding to an exposure time of about $381$~ks).

For the spectral analysis we only considered ScWs with \src\ within
$12^\circ$ 
from the centre of the 
(IBIS/ISGRI) 
field of view. 
We introduced this filter because
at larger off-axis angle the IBIS response is not 
well known\footnote{see the \inte\ data analysis documentation: \url{http://www.isdc.unige.ch/integral/analysis#Documentation}}.
We performed spectral analysis with \texttt{XSPEC} (ver. 12.7.0).
Systematic uncertainties of 1\% for IBIS/ISGRI and 3\% for JEM-X were added to our data sets
(according to the IBIS/ISGRI and JEM-X data analysis documentation). 

We obtained JEM-X light curves with a time resolution of $1$~s in the energy range $3-20$~keV.
For the timing analysis with IBIS/ISGRI data we used the \emph{ii\_light} 
routine\footnote{For a description of the tool \emph{ii\_light}, 
see \url{http://www.isdc.unige.ch/integral/download/osa/doc/10.0/osa_um_ibis/Cookbook.html}}
to obtain $18-50$ keV light curves with a time resolution 
of $1$~s. The bin times were corrected to the solar system barycentre.

\subsection{XMM-Newton} \label{sect. xmm}

\src\ has been observed with \xmm\ 
on 2006 October 12 (ObsID 0406140201)
with a single 19~ks exposure observation.

The data analysis was performed through the \xmm\
Science Analysis System (SAS) software (version 13.0.0).
We rejected time intervals affected by high background due to proton flares,
obtaining a total good exposure time of 14.4 ks.
We produced calibrated event lists for pn, MOS1, and MOS2
using the \texttt{EPPROC} (for pn) and \texttt{EMPROC} (for the two MOS cameras) tasks.
We reproduced the data reduction of \citet{Giunta09}, hence 
we considered for timing analysis only the pn event file,
because of the poor statistics in the MOS event files.
We extracted a $0.5-10$~keV source light curve
in a circular region with a radius of $20^{\prime\prime}$.
The background light curve was extracted in a source-free 
circular region with radius $50^{\prime\prime}$
of the same chip as the source.
We selected events with PATTERN$\leq$4 for the pn.
We corrected the times of the light curves to the 
solar system barycentre with the 
\texttt{BARYCEN} task, using the ephemeris JPL~DE405
(see e.g. \citealt{Hobbs06}).

\section{Timing analysis} 
\label{sect. timing analysis}

We analysed archival \inte\ observations 
with \src\ in outburst, during the period 2003--2010.
We report the detection of seven previously
unknown outbursts and eight outbursts previously observed
with \inte\ by \citet{Chernyakova03}, \citet{Molkov03}, 
\citet{Sguera07}, and \citet{Ducci10}.
With this we almost double the number of outbursts
detected with \inte.
Table \ref{table:list outbursts} shows the list
of 15 outbursts detected with \inte\ 
with a significance $> 5\sigma$
in the energy range $15-50$~keV.

Seven outbursts 
were observed with JEM-X.
The duration of the observations are of the order of hours to a few days.
Since \src\ was not always 
in the IBIS field of view during the outbursts,
we cannot constrain their total duration with our observations. 
A lower limit can be inferred from the observation date column of Table \ref{table:list outbursts}.

These observations show fast and transient flares, with durations
of a few hours. These flares are the brightest part ($L_{\rm x} \gtrsim 10^{35}$~erg~s$^{-1}$)
of a more complex flaring activity that characterizes SFXTs
and extends to $L_{\rm x} \approx 10^{33}$~erg~s$^{-1}$ \citep{Sidoli08}.
Because of the sensitivity limit of \inte,
we cannot study the flaring behaviour of \src\ as a whole,
which, on the other hand, has been studied by \citet{Romano10}
using the X-ray monitoring along an entire orbital period performed by \sw/XRT.

\begin{figure}
\includegraphics[bb=34 26 537 705,clip,angle=-90,width=9cm]{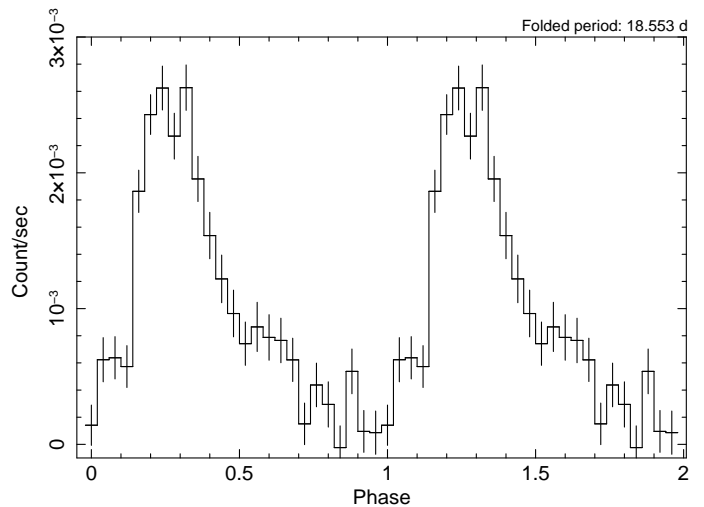}
\caption{\sw/BAT ($15-50$~keV) light curve of \src, folded at $P_{\rm orb}=18.553$~d
         and $T_{\rm epoch}=53415$ MJD.}
\label{fig:folded lcr}
\end{figure}

\subsection{Orbital X-ray modulation} \label{sect. orbital modulation}

We searched for the orbital period using the long-term
\sw/BAT light curve (averaged over the \sw\
orbital period of $\sim$90 minutes in the energy band $15-50$~keV)
of \src\ provided by the \sw/BAT team.
We retrieved the \sw/BAT orbital light curve
covering the data range from 2005 February 14
to 2013 May 6 (MJD range 53415$-$56418) from the
BAT Transient Monitor \citep[][]{Krimm2006_HEAD_BTM,Krimm2013:BATTM}
page\footnote{http://swift.gsfc.nasa.gov/docs/swift/results/transients/ }.
We excluded poor-quality points 
(quality flag 1, 2, and 3) and 
corrected the screened light curve for Earth's motion
using the \texttt{earth2sun} tool of the HEASARC software package 
\texttt{FTOOLS} v.6.12 \citep{Blackburn95}. We obtained a first estimate of the 
orbital period by applying an epoch-folding period search
(we used the \texttt{FTOOL} \texttt{efsearch}).
We refined the period measurement
by means of a phase-fitting technique 
(see e.g. \citealt{Dallosso03} for details).
The resulting orbital period is $18.553 \pm 0.003$ 
($1\sigma$ confidence level. $0.005$ at $2\sigma$ confidence level).
This measurement approaches the precision of the value 
determined by \citet{Levine11} using RXTE/ASM data,
and it is consistent with it within $2\sigma$ c.l.
The folded \sw/BAT light curve of \src\ 
is shown in Fig. \ref{fig:folded lcr}.

We also searched for the orbital period using the IBIS/ISGRI
light curve of \src\ ($17-80$~keV) provided by the online tool 
Heavens\footnote{http://www.isdc.unige.ch/integral/heavens}
of the INTEGRAL Science Data Centre. The light curve covers all
publicly available IBIS/ISGRI data of this source between 2003 March 6 to 2010 November 18
(MJD range 52704$-$55518). The resulting measurement of the orbital period of $18.57 \pm 0.02$ ($1\sigma$)
is consistent with the orbital period measurement obtained with the \sw/BAT data.
We used the measurement with \sw/BAT data of the orbital period and $T_{\rm epoch}=53415$~MJD
to calculate the orbital phases of the outbursts 
(see Table \ref{table:list outbursts}).

\begin{figure}
\includegraphics[bb=125 310 480 558,clip,width=9cm]{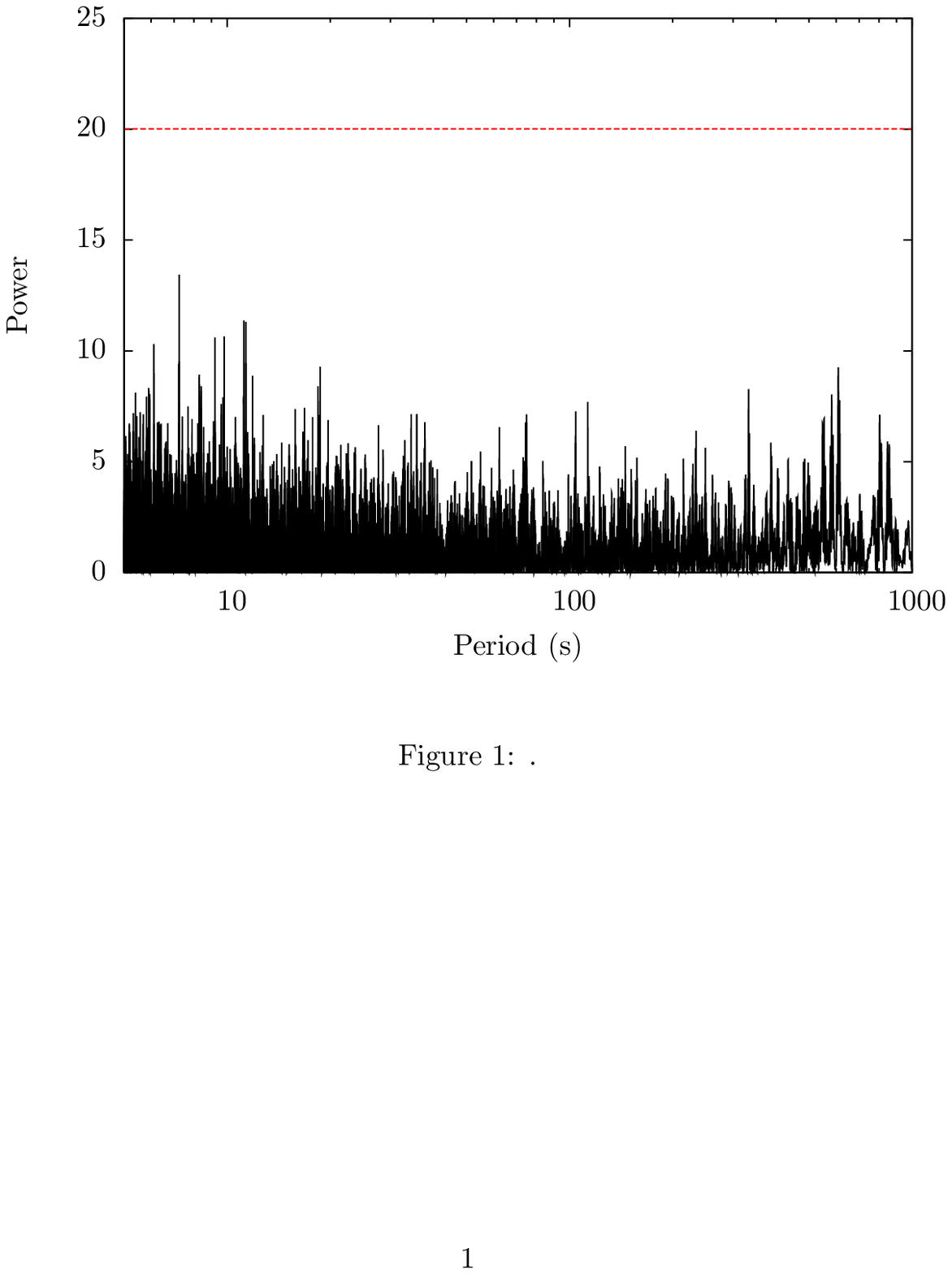}
\caption{Lomb-Scargle periodogram computed for JEM-X data of \src\ 
         during the outburst of 2006 April 19.}
\label{fig:lomb-scargle}
\end{figure}

\subsection{Searching for X-ray pulsations} \label{sect. timing integral}

\subsubsection{INTEGRAL} \label{subsect. timing integral}

We carried out a search for periodicity with both JEM-X and IBIS/ISGRI data
with the aim of monitoring the long-term evolution of the pulse period of \src.

For each new outburst (Table \ref{table:1}), we extracted JEM-X ($3-20$~keV) 
and IBIS/ISGRI ($18-50$~keV) light curves.
We applied the epoch-folding
and the fast Lomb-Scargle periodogram techniques (\citealt{Press89}; 
\citealt{Scargle82}; \citealt{Lomb76}).
The search for a timing feature was performed within 2~s and $5\times 10^3$~s.
For each time series, we estimated the number of independent frequencies
in the Lomb-Scargle periodograms using the formula (eq. 13) of \citet{Horne86}.

Since no statistically significant periodicities 
were detected in the new outbursts,
we performed a new timing analysis on the previously discovered 
outbursts of \src\ (Table \ref{table:1}).
In particular, we focused our attention on  
outburst No.\,8, where a periodic signal 
at $\sim$21~s was detected by \citet{Sguera07} in the JEM-X light curve.
We searched the $\sim$21~s signal in both JEM-X and IBIS/ISGRI light curves,
but no statistically significant periodicity was detected.

We also performed a timing analysis on the JEM-X light curve by
varying the bin time from $0.5$ to $2$~s
and for three different energy ranges: $3-20$~keV, 
$4-20$~keV (the energy band adopted by \citealt{Sguera07}), and $3-35$~keV.

Fig. \ref{fig:lomb-scargle} shows the Lomb-Scargle periodogram obtained with
the $4-20$~keV JEM-X light curve (bin time 1~s) of outburst No.\,8.
The dashed red line corresponds to a false-alarm probability of 0.1\%.
We performed Monte Carlo simulations on this observation to set a 3$\sigma$
upper limit on the pulsed fraction between 15 and 25~s.
We found an upper limit of $\sim$28\% in the $4-20$~keV JEM-X light curve.

For each outburst, we also searched for periodicities on very short time-scales,
up to milliseconds, in the IBIS/ISGRI and JEM-X event arrival times,
both corrected to the solar system barycentre.
To increase the signal-to-noise ratio we selected only events
with a pixel illuminated fraction (PIF, pixel fraction illuminated 
by the source) equal to 1. 
Again, no statistically significant periodicity was detected.

We also repeated the data analysis for outburst No. 8
with the previous versions of the analysis software
available to us (OSA 6.0 to 9.0).
The results are compatible with those obtained using OSA 10.0.
Since OSA 5.1, used by \citet{Sguera07}, was not
available to us, we cannot completely reproduce their data analysis,
hence we cannot establish the origin of the difference in results.

\begin{figure}
\includegraphics[bb=125 310 480 558,clip,width=9cm]{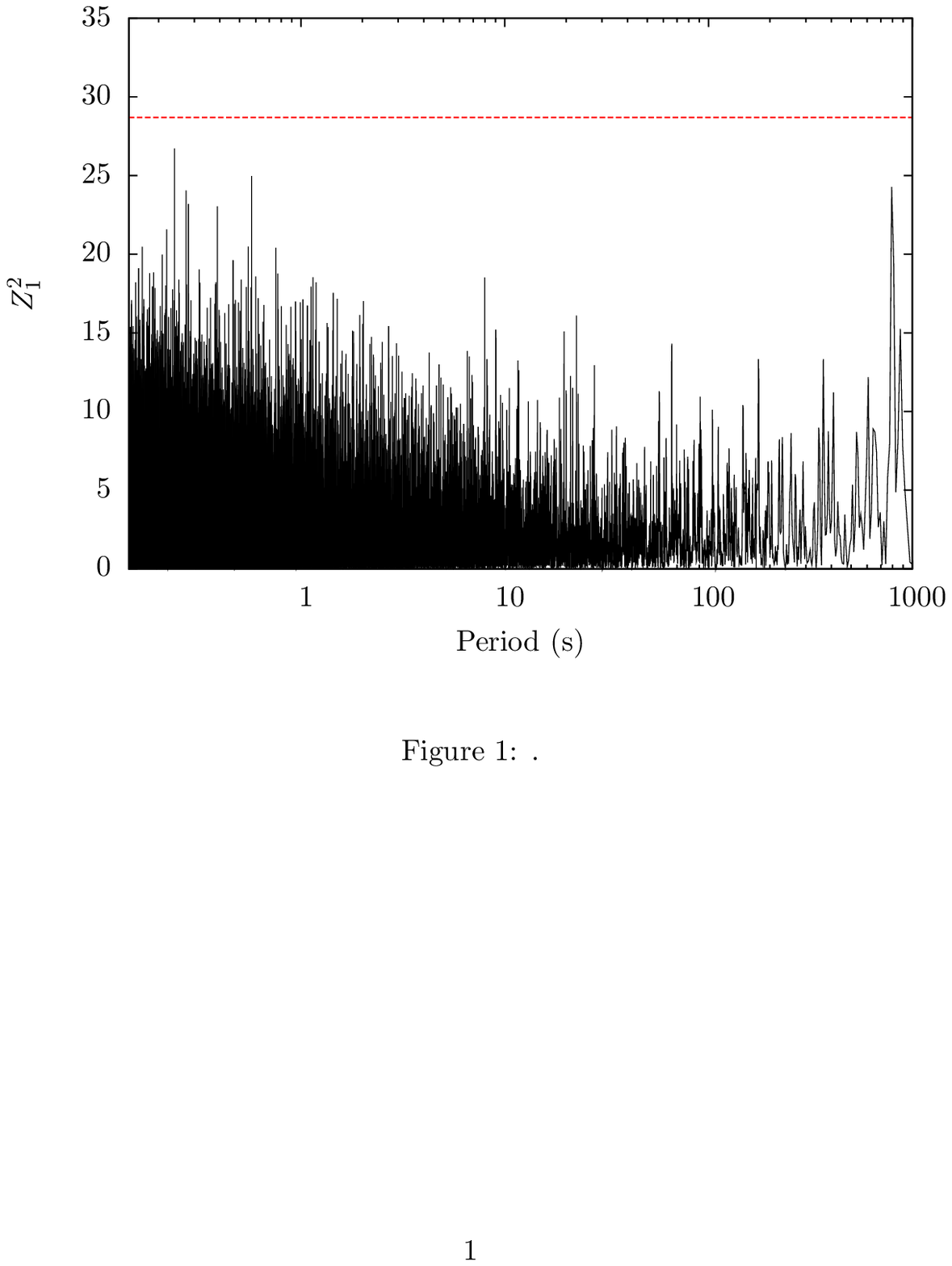}
\caption{Rayleigh test ($Z^2_1$) computed for the \xmm/pn data of \src. The dashed red line corresponds 
to a 5$\sigma$ detection.}
\label{Z-xmm}
\end{figure}

\subsubsection{XMM-Newton}

After extracting the pn event files, 
we searched for a periodic modulation in the 
$0.5-10$ keV barycentred event arrival times of \src.
The periodicity search was limited by the $0.073$~s time resolution
of pn (in full-frame mode) and the total duration of the observation ($14.4$~ks).

We performed the H-test \citep{deJager89}
in the frequency range $\Delta f = 0.00014-7$~Hz.
No significant features appeared in the H-test periodogram.
We confirm the presence of a peak in the periodogram at frequency $\sim$0.0475~Hz.
By applying the $Z^2_n$ test for 1 to 4 harmonics
we found at $f \sim$0.0475~Hz
a significant contribution from the second harmonic
($Z^2_1=12.5$, $Z^2_2=23.7$). 
Fig. \ref{Z-xmm} shows the $Z^2_1$ periodogram obtained from 
the $0.5-10$~keV  pn arrival times of \src.

The probability to obtain a noise peak
of $Z^2_2=23.7$ in a single trial is $p=\exp{(-Z^2_2/2)}=7.1 \times 10^{-6}$,
which corresponds to a detection of pulsations at a confidence level of $99.9993$\%.
The probability to obtain that noise peak in $N=4T_{\rm exp} \Delta f$ 
independent trials is $p=1 -[1 -\exp{(-Z^2_2/2)}]^N= 0.94$
(where 4 is the number of harmonics at which the $Z^2_n$ test has been performed), 
which corresponds to a detection of a pulsation at a confidence level of $6$\%.
Therefore, the signal at $\sim$21~s 
is most likely due to a statistical fluctuation.
We used the highest observed peak in the $Z^2_1$ periodogram and Monte
Carlo simulations to set a limit to a true signal. The 3$\sigma$ upper
limit on the pulsed fraction of a sinusoidal signal is $\sim$30\% in
the 0.5--10 keV band and for periods between 15~s and 25~s.
We point out that our results are compatible with those of \citet{Giunta09}.
The main difference lies in the size of the window of frequencies where the search was carried out.
On the basis of the periodicity reported by \citet{Sguera07},
\citet{Giunta09} searched for a periodicity in a small window 
centred on $21.0526$~s.
In light of the new results obtained in Sect. \ref{subsect. timing integral} 
from the analysis of the JEM-X data of the outburst No. 8, we selected a wider window 
of frequencies, which reduces the significance of the peak reported by \citet{Giunta09}.

\begin{table*}
\caption{Spectral parameters of the outbursts of \src\ observed with \inte\ and summarized in Table \ref{table:list outbursts}. 
For each outburst, we report the peak flux observed with IBIS/ISGRI and
the best-fitting parameters obtained from IBIS/ISGRI, JEM-X, and IBIS/ISGRI+JEM-X spectral analyses. 
We obtained the best fit using an absorbed power-law or a cutoff-power-law model.
Listed uncertainties are at the 90\% confidence level.}             
\label{table:1}      
\centering   
\begin{tabular}{lcccccccc} 
\hline\hline  
        \noalign{\smallskip}   
No.    &    $18-50$~keV peak flux      & \multicolumn{3}{c}{ISGRI spectral parameters} & \multicolumn{4}{c}{JEM-X\,$+$\,ISGRI spectral parameters}   \\
       &$10^{-10}$~erg~cm$^{-2}$~s$^{-1}$ &        $\Gamma$        &    $E_{\rm c}$ (keV)      & $\chi^2_\nu$ (d.o.f.) & $N_{\rm H}$ ($10^{22}$ cm$^{-2}$) & $\Gamma$ & $E_{\rm c}$ (keV) & $\chi^2_\nu$ (d.o.f.) \\ 
\noalign{\smallskip}
\hline                        
\noalign{\smallskip}
{ }\,1 &        $2.0 \pm 0.4$    & $2.3{+0.6 \atop -0.6}$ &                    & $0.881$ $(12)$ & $12{+17 \atop -11}$  & $2.2{+0.6 \atop -0.6}$ &                         & $0.869$ $(20)$ \\ 
\noalign{\smallskip}
{ }\,2 &        $3.1 \pm 0.5$    & $2.6{+0.6 \atop -0.6}$ &                    & $1.124$ $(14)$ &                      &                        &                         &                \\
\noalign{\smallskip}
{ }\,3 &        $5.5 \pm 0.5$    & $2.5{+0.2 \atop -0.1}$ &                    & $1.343$ $(14)$ &                      &                        &                         &                \\
\noalign{\smallskip}
{ }\,4 &        $4.6 \pm 0.4$    & $2.5{+0.2 \atop -0.2}$ &                    & $1.291$ $(16)$ &                      &                        &                         &                \\
\noalign{\smallskip}
{ }\,5 &        $5.8 \pm 0.4$    & $0.9{+0.6 \atop -0.9}$ & $21{+13 \atop -8}$ & $1.775$ $(14)$ &                      &                        &                         &                \\
\noalign{\smallskip}
{ }\,6 &        $5.8 \pm 0.5$    & $2.7{+0.3 \atop -0.3}$ &                    & $1.285$ $(12)$ &                      &                        &                         &                \\
\noalign{\smallskip}
{ }\,7 &        $4.2 \pm 0.5$    & $2.4{+0.3 \atop -0.3}$ &                    & $1.191$ $(13)$ &                      &                        &                         &                \\
\noalign{\smallskip}
{ }\,8 &        $8.3 \pm 0.4$    & $1.4{+0.5 \atop -0.5}$ & $30{+24 \atop -10}$& $1.100$ $(14)$ &$5.1{+4.1 \atop -3.6}$& $1.3{+0.3 \atop -0.3}$ &    $27{+8 \atop -5}$    & $0.856$ $(26)$ \\ 
\noalign{\smallskip}
{ }\,9 &        $7.6 \pm 1.0$    & $2.7{+0.2 \atop -0.2}$ &                    & $1.750$ $(10)$ &                      &                        &                         &                \\
\noalign{\smallskip} 
10     &        $8.6 \pm 0.7$    & $2.5{+0.2 \atop -0.2}$ &                    & $1.485$ $(15)$ &   $19{+9 \atop -8}$  & $2.6{+0.2 \atop -0.2}$ &                         & $1.044$ $(26)$ \\ 
\noalign{\smallskip}
11     &        $4.0 \pm 0.6$    & $2.2{+0.3 \atop -0.2}$ &                    & $1.106$ $(13)$ &   $9{+7 \atop -6}$   & $2.2{+0.2 \atop -0.2}$ &                         & $0.824$ $(24)$ \\ 
\noalign{\smallskip}
12     &        $6.7 \pm 0.6$    & $2.2{+0.2 \atop -0.2}$ &                    & $1.642$ $(13)$ &                      &                         &                        &                \\
\noalign{\smallskip}
13     &        $6.8 \pm 0.8$    & $2.6{+0.3 \atop -0.3}$ &                    & $1.891$ $(12)$ &   $28{+10 \atop -8}$ & $2.5{+0.3 \atop -0.3}$  &                        & $1.201$ $(24)$ \\ 
\noalign{\smallskip}
14     &        $4.2 \pm 0.6$    & $2.9{+0.4 \atop -0.4}$ &                    & $0.947$ $(13)$ &  $60{+24 \atop -19}$ & $2.8{+0.4 \atop -0.3}$  &                        & $0.923$ $(25)$ \\ 
\noalign{\smallskip}
15     &        $4.9 \pm 0.7$    & $2.8{+0.3 \atop -0.3}$ &                    & $1.808$ $(13)$ &  $34{+13 \atop -10}$ & $2.7{+0.3 \atop -0.3}$  &                        & $1.400$ $(24)$ \\ 
\noalign{\smallskip}
\hline                         
\end{tabular}
\end{table*}

\section{Spectral analysis} 
\label{sect. spectral analysis}

We fitted the IBIS/ISGRI spectra of the newly discovered outbursts
of Table \ref{table:list outbursts} with several spectral models.
We also repeated the spectral analysis of outbursts Nos.\,1-8
to benefit from the new energy calibration of IBIS/ISGRI
that has been implemented in OSA 10 (see e.g. \citealt{Caballero13}).
We also simultaneously fitted IBIS/ISGRI and JEM-X spectra
of the outbursts for which JEM-X data were available.
We included constant factors in the spectral fitting to allow 
for normalization uncertainties between the instruments.

For each outburst, we obtained good fits with power-law or cutoff power-law 
models (see Table \ref{table:1}).
Outbursts Nos.\,5 and 8 have the longest exposure time.
Their spectra are best reproduced with an absorbed cutoff power-law model.
A power-law (absorbed when JEM-X data were available) is an adequate description
of the observed spectrum for the remaining 13 outbursts,
which have the poorest quality data sets.
We point out that these spectra can be well fitted  by a simple absorbed power-law
because of their poor statistics coupled with the high e-folding energy
(whose value is inversely proportional to the slope of the spectrum at energies
above the energy cutoff) that we obtained from the average fit of the spectrum
of \src\ with a high-energy cutoff component (\texttt{highecut} in \texttt{XSPEC}, see below).
The spectral parameters we obtained agree with previous studies
of the hard X-ray spectrum of \src\ (see Sect. \ref{sect. introduction}).

We derived column densities in excess of the one along the line of sight
($N_{\rm H} \approx 1.6 \times 10^{22}$~cm$^{-2}$; \citealt{Kalberla05}).
The measured $N_{\rm H}$ varies from $N_{\rm H} \approx 5.1 \times 10^{22}$~cm$^{-2}$ (outburst No.\,8)
to $N_{\rm H} \approx 6.0  \times 10^{23}$~cm$^{-2}$ (outburst No.\,14).
However, all of the measured values of $N_{\rm H}$ between observations are compatible
within their associated errors.
Similarly high and variable column densities have been observed in many \inte\ HMXBs (e.g. \citealt{Tomsick09}). 
For \src, there is no evidence (here and in previous works) 
for a correlation of the column density with the source flux (see Table \ref{table:1}).

We did not find evidence of variability 
in the X-ray spectrum between different outbursts 
and for different luminosities.
Therefore, we extracted the average JEM-X1, JEM-X2, and IBIS/ISGRI spectra
and fitted them
with several phenomenological models.

A simple absorbed power-law gave an unacceptable fit 
with a high $\chi^2 = 212/101$,
caused mostly by strong residuals at the lowest (below $\sim$5~keV)
and highest energies (above $\sim$50~keV).
Replacing the power-law with a cutoff power-law
or a power-law modified at energies above the 
energy cutoff $E_{\rm c}$ by $\exp[(E_{\rm c} - E)/E_{\rm f}]$,
(where $E_{\rm f}$ is called e-folding energy)
improved the fit (see Table \ref{table:spec tot}).
\begin{figure}
\includegraphics[bb=77 0 585 730,clip,angle=-90,width=9cm]{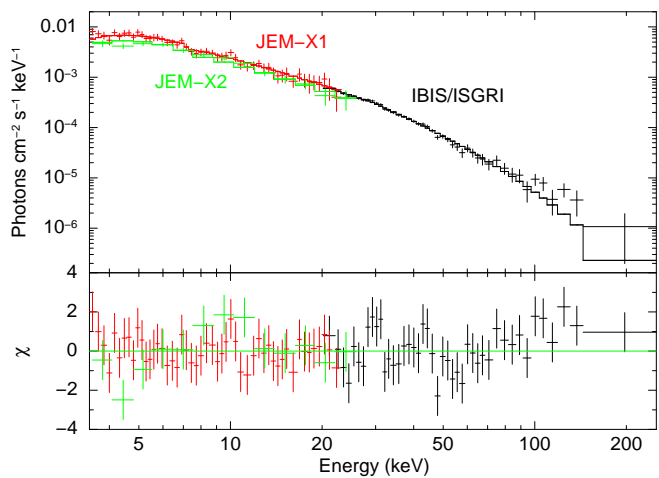}
\caption{Joint JEM-X1 (red), JEM-X2 (green), IBIS/ISGRI (black) spectrum fitted with a high-energy cutoff power-law,
         with residuals in units of standard deviations. 
         The parameters of the best fit are summarized in Table \ref{table:spec tot}.}
\label{fig:spec}
\end{figure}
\begin{table}
\caption{Best-fitting parameters of the absorbed cutoff power-law,
         high-energy cutoff power-law,
         \texttt{COMPTT}, and \texttt{COMPMAG} models
         for the joint JEM-X plus IBIS/ISGRI spectra
         (see Figs. \ref{fig:spec} and \ref{fig:spec2}).
         For \texttt{COMPTT} model $kT_{\rm seed}$ corresponds to $kT_{\rm w}$.
         For \texttt{COMPMAG} model $kT_{\rm seed}$ corresponds to $kT_{\rm bb}$. 
         We assumed a free-fall velocity profile for the accreting matter
         ($\eta=0.5$), $\beta_0=0.05$, $r_0=0.25$, $A=1$.
         Listed uncertainties are at the 90\% confidence level.}
\label{table:spec tot}      
\centering   
\begin{tabular}{lcc} 
\hline\hline   
       \noalign{\smallskip}     
Parameter                      &     cutoff power-law     &  power-law with            \\
                               &                         &  high-energy cutoff       \\
  \noalign{\smallskip}
\hline 
  \noalign{\smallskip}
$N_{\rm H}$ ($10^{22}$ cm$^{-2}$) & $9.3{+3.3 \atop -3.1}$  &     $13{+3 \atop -3}$     \\
\noalign{\smallskip}
$\Gamma$                       & $1.5{+0.2 \atop -0.2}$  & $1.97{+0.13 \atop -0.14}$ \\
\noalign{\smallskip}
$E_{\rm c}$ (keV)               &   $31{+7 \atop -5}$     &     $29{+2 \atop -2}$     \\
\noalign{\smallskip}
$E_{\rm f}$ (keV)               &                         &     $42{+8 \atop -6}$     \\
\noalign{\smallskip}
$\chi^2_\nu$ (d.o.f.)          &     $1.011$ ($101$)      &     $0.891$ ($100$)      \\
  \noalign{\smallskip}
\hline\hline   
\noalign{\smallskip}     
Parameter                      &      \texttt{COMPTT}      &       \texttt{COMPMAG}            \\
\noalign{\smallskip}
\hline 
\noalign{\smallskip}
$N_{\rm H}$ ($10^{22}$ cm$^{-2}$) &    $14{+3 \atop -3}$      &    $10.5{+1.8 \atop -1.7}$        \\
\noalign{\smallskip}
$kT_{\rm seed}$ (keV)            &  $0.4{+0.4 \atop -0.4}$   &     $0.3{+0.2 \atop -0.3}$        \\
\noalign{\smallskip}
$kT_{\rm e}$ (keV)              &    $14{+3 \atop -2}$      &       $23{+18 \atop -12}$          \\
\noalign{\smallskip}
$\tau$                         &  $5.0{+0.9 \atop -0.9}$   &   $0.226{+0.003 \atop -0.005}$     \\
\noalign{\smallskip}
normalization                  & $0.03{+1.03 \atop -0.03}$ & $2.6{+9.8 \atop -1.5} \times 10^4$ \\
\noalign{\smallskip}
$\chi_\nu^2$ (d.o.f.)          &       $1.195$ (100)        &         $1.009$ (100)              \\
\noalign{\smallskip}
\hline                                    
\end{tabular}
\end{table}
The fit with a simple cutoff power-law has a better $\chi^2_\nu$
compared with the high-energy cutoff power-law model,
which overfits the data with a $\chi^2_\nu$ of $0.891$.
The high-energy cutoff power-law model is typically used 
to describe the X-ray emission from accreting
compact objects in HMXBs.
We obtained $\Gamma \simeq1.97$, $E_{\rm c}\simeq29$~keV,
and $E_{\rm f}\simeq42$~keV (see Fig. \ref{fig:spec} and Table \ref{table:spec tot}).
The energy cutoff and the e-folding energy obtained
from the fit show that the X-ray spectrum of \src\
is harder than the X-ray spectra of typical HMXBs.
Indeed, $E_{\rm c}$ typically ranges from $\approx7$~keV
to $\approx25$~keV, and $E_{\rm f}$ from $\approx5$~keV
to $\approx40$~keV (see e.g. \citealt{Coburn02}; \citealt{White83}).
Although the $\chi^2$ and residuals of the fit with 
an absorbed power-law with high-energy cutoff are acceptable,
a positive trend above $\sim$50~keV, 
ascribable to a poorly modelled hard excess,
is present in the residuals.

\begin{figure}
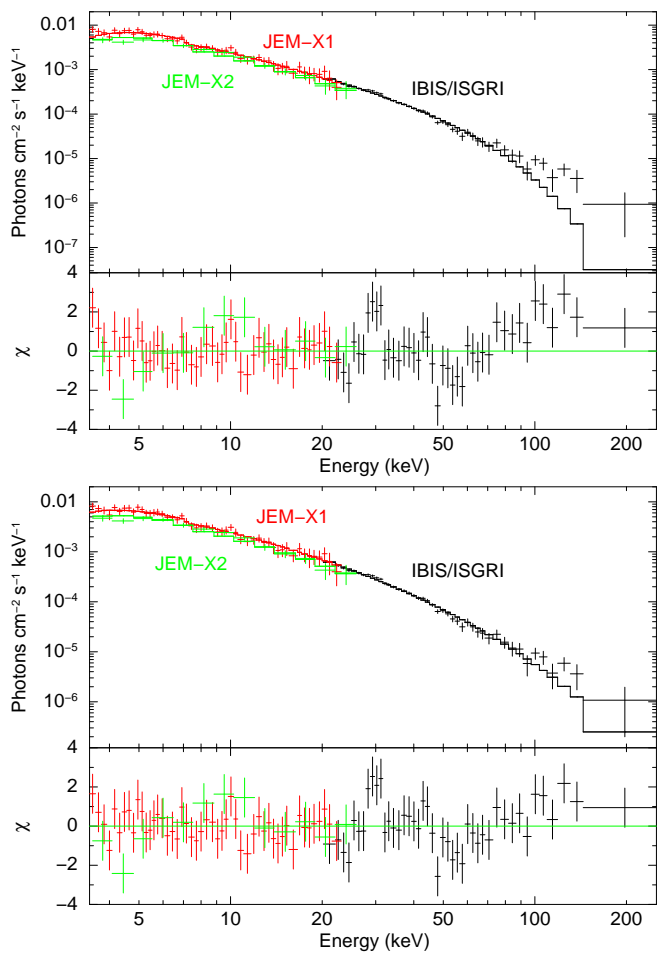

\includegraphics[bb=77 0 585 730,clip,angle=-90,width=9cm]{figure5a.ps}
\includegraphics[bb=77 0 585 730,clip,angle=-90,width=9cm]{figure5b.ps}
\caption{Joint JEM-X1 (red), JEM-X2 (green), IBIS/ISGRI (black) spectrum fitted with an absorbed \texttt{COMPTT} model (upper panel)
         and an absorbed \texttt{COMPMAG} model (lower panel). 
         The best-fit parameters are summarized in Table \ref{table:spec tot}.}
\label{fig:spec2}
\end{figure}

To better reproduce the shape of the observed spectrum 
we also fitted the spectrum with the physical models
\texttt{COMPTT} \citep{Titarchuk94} and \texttt{COMPMAG} \citep{Farinelli12}.

\texttt{COMPTT} is an analytic model developed by \citet{Titarchuk94},
which can reproduce the spectrum emerging from 
the Comptonization of soft photons with temperature $T_{\rm w}$
by a hot plasma of electrons with temperature $T_{\rm e}$
and optical depth $\tau$.
The input spectrum for the seed photons is a Wien law.
The origin of the seed photons is often associated 
with the accretion disc or the neutron star.

\texttt{COMPMAG} \citep{Farinelli12} is designed to reproduce the spectrum
of a magnetized accreting pulsar,
in the case of cylindrical accretion on to the polar cap of the neutron star.
The authors considered a blackbody spectrum of seed photons
Comptonized by the plasma. Both thermal and bulk Comptonization were considered.
The parameters of the model are the temperature $T_{\rm bb}$ of the 
blackbody seed photons, the electron temperature $T_{\rm e}$,
the vertical optical depth $\tau$ of the comptonization plasma,
the radius of the accretion column $r_0$, and the albedo at the star surface $A$.
The blackbody photons are produced by the accreted matter throughout the accretion column.
The \texttt{COMPMAG} model also allows one to select the velocity profile $\beta(Z)$
of the infalling material in the accretion column between two options:
a velocity increasing towards the neutron star surface $\beta(Z) \propto Z^{-\eta}$ (where $\eta$ is also
a parameter that controls the steepness of the vertical velocity profile, 
and $Z$ is the altitude above the surface of the neutron star)
or, following \citet{Becker07}, a decelarating profile $\beta(\tau) \propto -\tau$.
If the first option is selected, $\eta$ and the parameter $\beta_0$,
which governs the terminal velocity of the accreting matter at the neutron star surface, 
are also introduced as spectral parameters.

Following the procedure applied by \citet{Farinelli12_sfxts}
to the \sw\ spectra of the SFXTs XTE~J1739$-$302 and IGR~J17544$-$2619,
we assumed an increasing velocity of the infalling material 
towards the neutron star surface and set
$\eta=0.5$ and $\beta_0=0.2$ or $0.05$.
We also set $r_0=0.25$ (in units of the neutron star
Schwarzschild radius) and $A=1$.

We obtained acceptable fits
with either \texttt{COMPTT} (with both spherical and disc geometry)
or \texttt{COMPMAG} models
(Fig. \ref{fig:spec2}; Table \ref{table:spec tot}).
The spectral fits with the absorbed \texttt{COMPTT}
(Fig. \ref{fig:spec2}) shows a hard excess at energies above $\sim$80~keV.
To a lesser extent, the absorbed \texttt{COMPMAG} also shows a positive trend in the residuals
that could indicate a hard excess above $\sim$80~keV.
We noticed that a positive trend in the residuals above $\sim$50~keV 
was also present in the spectral fits of the
outbursts with the highest quality data sets (outbursts 5 and 8).
To determine whether the hard excess observed in the average spectrum
was not solely due to the data sets of outbursts 5 and 8,
we produced an average JEM-X and IBIS/ISGRI spectrum without 
these two outbursts. We found that the hard excess
was still present in the spectral fit.
We tested for a hard excess above  $\sim$80~keV
using the F-test.
We added a power-law with a pegged normalization (\texttt{pegpwrlw} in \texttt{XSPEC})
component to the absorbed \texttt{COMPMAG} spectral model
and obtained an acceptable fit with $\chi^2_\nu = 0.96$ (98 d.o.f.)
without the hard excess in the residuals.
The photon index of the power-law is $0.1\pm 1.0$,
and we set the energy ranges of the power-law to $80-250$~keV.
The F-test gives a $3.3$\% probability of a chance improvement of the $\chi^2_\nu$,
which is not significant at the 3$\sigma$ level.

We compared the spectral parameters resulting from the fit
of the \src\ spectrum with \texttt{COMPMAG} with previously reported
uses of this model. With the exception of the electron temperature $T_{\rm e}$,
the best-fit parameters of \src\ agree with the values from the joint
XRT/BAT spectra of the SFXTs XTE~J1739$-$302 and IGR~J17544$-$2619 \citep{Farinelli12_sfxts}.
The obtained value of $kT_{\rm e}$ in \src\ ($\sim$23~keV)
is significantly higher than those obtained by  \citet{Farinelli12_sfxts}
for XTE~J1739-302 and IGR~J17544-2619
($9.62{+4.05 \atop -2.57}$~keV, $3.47{+0.34 \atop -0.02}$~keV respectively).
The high electron temperature resulting from the fit of the \src\ spectrum
is appropriate to account for the relatively hard X-ray emission observed by \inte.

The hard X--ray spectrum of \src\ during outbursts is not typical of HMXBs.
However, the HMXB X Persei shows an extremely hard X-ray spectrum
(\citealt{Doroshenko12} and references therein).
It is a persistent accreting pulsar with a spin period of about 
$837$~s, which orbits around a Be star in $\sim$250 days 
in an eccentric orbit ($e \sim0.11$).
Recently, \citealt{Doroshenko12} extracted an average spectrum 
using data from ISGRI, the spectrometer on \inte\ (SPI; \citealt{Vedrenne03}), 
and JEM-X1 instruments ($t_{\rm exp} \sim400$~ks).
They modelled the $4-200$~keV \inte\ spectrum of X~Persei
with a model with two Comptonization components with independent
electron temperatures and optical depths, where the component
at lower energies describes the thermal Comptonization
and the component at higher energies describes the bulk Comptonization
in the accretion flow of the soft photons emerging from the polar cap.
\begin{figure}
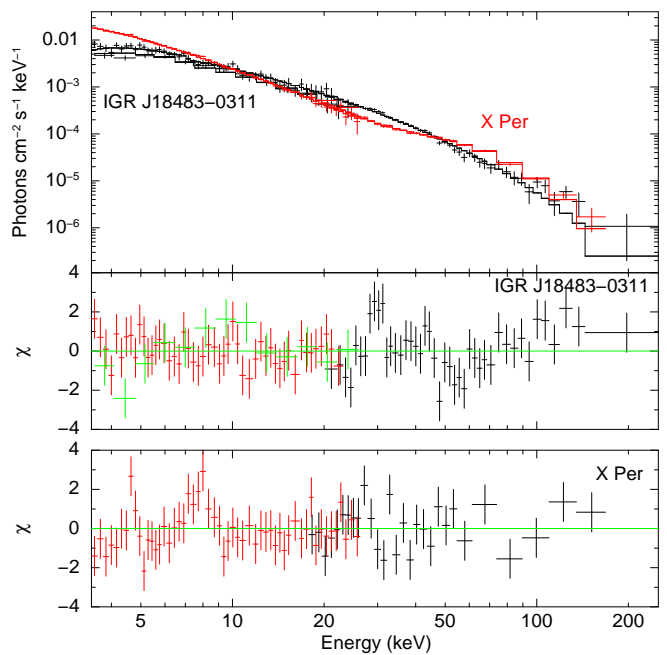

\includegraphics[bb=77 0 537 730,clip,angle=-90,width=9cm]{figure6a.ps}
\includegraphics[bb=349 0 585 730,clip,angle=-90,width=9cm]{figure6b.ps}
\caption{Comparison of the spectra of \src\ (black line) fitted with an absorbed \texttt{COMPMAG} model
         (the spectral parameters are summarized in Table \ref{table:spec tot}) 
         and X~Persei (red line) fitted with two Comptonization models 
         (see details in \citealt{Doroshenko12}).}
\label{fig:spec3}
\end{figure}
A comparison between the spectrum
of \src\ modelled with \texttt{COMPMAG} and the spectrum of X Persei
modelled with two Comptonization components
(see \citet{Doroshenko12} for details) confirms
that both sources exhibit hard spectra
with unusually high cutoff and e-folding energies
(Fig. \ref{fig:spec3}).

\section{Conclusions}
\label{sect. conclusions}

We analysed archival \inte\ data of the SFXT \src,
covering the data range 2003--2010
(corresponding to an exposure time of $\sim$381~ks).
We detected 15 outbursts, seven of which are reported for the first time.

A timing analysis performed on the IBIS/ISGRI, JEM-X,
and \xmm\ light curves of \src\ did not reveal
a significant periodicity.
In particular, we did not find the pulsation at $\sim$21~s
in the JEM-X and \xmm\ observations where it was previously detected
by \citet{Sguera07} and \citet{Giunta09}.
In light of the results reported here and in previous works 
(see Sect. \ref{sect. introduction}), 
the lack of detecting a periodic signal attributable
to the spin period of a pulsar
may be due to the low statistics
of the available data sets,
a weak magnetic field of the neutron star, 
or an alignment of the spin and magnetic-field axes.
Alternatively, the compact object may be a black hole.
We point out that there are no confirmed 
black holes in SFXTs to date. Therefore, such an identification
would be particularly important.

We also performed spectral studies on each outburst
and on the average broadband spectrum.
For the first time we applied physical models based on thermal
and bulk Comptonization processes to describe 
the hard X-ray emission of \src.
We obtained a good fit
when modelling the X-ray emission with an absorbed \texttt{COMPMAG} model.
We also obtained acceptable fits with the phenomenological models
of a cutoff power-law or a power-law with high-energy cutoff. The relatively high values of
$E_{\rm c}$, $E_{\rm f}$, and $kT_{\rm e}$
obtained from these fits suggest that the spectrum of \src\
is harder than the X-ray spectra of typical HMXBs.
The spectral residuals of Figs. \ref{fig:spec} and \ref{fig:spec2}
show a positive trend above $\sim$80~keV that could indicate a hard excess.
However, when we added a power-law with pegged normalization
component to the absorbed \texttt{COMPMAG} spectral model,
the F-test gave a $3.3$\% probability of a chance improvement of the $\chi^2_\nu$,
which is not significant at the 3$\sigma$ level (Sect. \ref{sect. spectral analysis}).
More sensitive observations may be able to verify whether there is such an excess.

\begin{acknowledgements}
We thank the anonymous referee for his/her useful and constructive comments,
which helped to improve the paper.
This paper is based on data from observations with INTEGRAL, XMM-Newton, and Swift.
INTEGRAL is an ESA project with instruments and science data centre funded by ESA
member states (especially the PI countries: Denmark, France, Germany,
Italy, Spain, and Switzerland), Czech Republic and Poland,
and with the participation of Russia and the USA.
XMM-Newton is an ESA science mission with instruments and contributions directly
funded by ESA Member States and NASA.
This work made use of the results of the Swift/BAT hard X-ray transient monitor:
http://swift.gsfc.nasa.gov/docs/swift/results/transients/ .
This research is funded by the Deutsche Forschungsgemeinschaft
through the Emmy Noether Research Grant SA 2131/1.
L.D. thanks Vito Sguera for his helpful advice. 
\end{acknowledgements}

\bibliographystyle{aa} 
\bibliography{igrj18483}

\end{document}